\documentclass[]{aa}  
\usepackage{graphicx}
\usepackage[sort]{natbib}
\usepackage{txfonts}
\newcommand{\uJy}{\ensuremath{\mu{\rm Jy}}}

\newcommand{\um}{\ensuremath{\mu{\rm m}}}
\begin{document}
   \title{The first VLBI image of an Infrared-Faint Radio Source}

   \subtitle{}

   \author{E. Middelberg\inst{1}
           \and
           R. P. Norris\inst{2}
           \and
           S. Tingay\inst{3}
           \and
           M. Y. Mao\inst{4}
           \and
           C. J. Phillips\inst{2}
           \and
           A. W. Hotan\inst{3}
          }

   \institute{Astronomisches Institut, Ruhr-Universit\"at Bochum, Universit\"atsstr. 150,
              44801 Bochum, Germany\\
              \email{middelberg@astro.rub.de}
         \and
             Australia Telescope National Facility, PO Box 76, Epping NSW 1710, Australia\\
             \email{ray.norris@csiro.au, chris.phillips@csiro.au}
         \and
             Department of Imaging and Applied Physics, Curtin University of Technology,
	     Bentley, Western Australia, Australia\\
	     \email{s.tingay@ivec.org, a.hotan@curtin.edu.au}
         \and
             School of Mathematics and Physics, University of Tasmania, Private Bag 21,
             Hobart TAS 7001, Australia\\
             \email{minnie.mao@utas.edu.au}
             }

   \date{Received, accepted}

 
  \abstract
   {To investigate the joint evolution of active galactic nuclei and
   star formation in the Universe.}
   {In the 1.4\,GHz survey with the Australia Telescope Compact Array
   of the Chandra Deep Field South and the European Large Area ISO
   Survey\,-\,S1 we have identified a class of objects which are
   strong in the radio but have no detectable infrared and optical
   counterparts. This class has been called Infrared-Faint Radio
   Sources, or IFRS. 53 sources out of 2002 have  been
   classified as IFRS. It is not known what these objects are.}
   {To address the many possible explanations as to what the nature of
   these objects is we have observed four sources with the Australian
   Long Baseline Array.}
   {We have detected and imaged one of the four sources
   observed. Assuming that the source is at a high redshift, we find
   its properties in agreement with properties of
   Compact Steep Spectrum sources. However, due to the lack of optical
   and infrared data the constraints are not particularly strong.}
   {}


   \maketitle
%

\section{Introduction}

Infrared-Faint Radio Sources (IFRS) were recently discovered as a
class by \cite{Norris2006a}, and may be related to the Optically
Invisible Radio Sources (OIRS) identified by
\cite{Higdon2005}. IFRS are radio sources which have no
counterparts in infrared images from the {\it Spitzer} Wide-Area
Extragalactic Survey (SWIRE) between 3.6\,\um\ and 24\,\um, and are
discovered in arcsec-scale radio observations. They are unexpected
because it was thought that any galaxy which is detected in radio
observations should be detected in the infrared with relatively short
integrations. Assuming the SED of known classes of galaxy, a 5\,mJy
radio source in the local Universe should produce a detectable Spitzer
source, regardless of whether it is generated by star formation or AGN
(active galactic nuclei). Similarly a normal L$_*$ galaxy at $z<1$,
whether spiral or elliptical, should be visible in our Spitzer and
I-band data.

\cite{Norris2006a} and \cite{Middelberg2008a} together identified 53
such sources out of 2002 (2.7\,\%) detected in the ATLAS survey,
co-located with the SWIRE survey. Most of these sources have flux
densities of only a few hundred microjansky, but some are strong and
have flux densities of more than 20\,mJy. Stacking 3.6\,\um\ {\it
Spitzer} images at the positions of 22 IFRS, \cite{Norris2006a} were
unable to make a detection in the averaged image and so demonstrated
that IFRS are well below the detection threshold of the SWIRE survey.

The nature of IFRS and the reason for their faintness at infrared
wavelengths is unclear. Possible explanations are that (i) these
sources are extremely redshifted Active Galactic Nuclei (AGN); (ii)
they are dust-rich, extremely obscured galaxies which makes them
invisible in the infrared; (iii) they are lobes of nearby,
unidentified radio galaxies; or (iv) they are an unknown type of
galactic or extragalactic object. Because IFRS have so far only been
detected at radio wavelengths it is not possible to measure their
redshifts, as spectroscopy requires sub-arcsecond positional accuracy,
which the radio observations cannot provide. Also, the comparatively
low resolution of the radio images makes it difficult to select the
correct optical counterpart, because the corresponding optical
observations are deep, and hence confusion-limited.

A promising route to find out more about IFRS is radio observations
with Very Long Baseline Interferometry (VLBI). VLBI observations are
sensitive only to very compact structures with brightness temperatures
of the order of $10^6$\,K or more, which are unambiguous signposts of
AGN activity. They also yield, when astrometric calibrators are used,
positions accurate to milliarcseconds, and milliarcsecond-scale
morphologies which can be interpreted in terms of the emission
mechanism at work.

\cite{Norris2007b} have observed two IFRS with VLBI and discovered
one. Unfortunately, their $(u,v)$ coverage was too poor to make a
reliable image of the detected source. They concluded that the VLBI
observations were consistent with a radio-loud AGN at high redshift,
or with a lower-power AGN at lower redshift in an abnormally obscured
galaxy.

\begin{figure}
\centering
\includegraphics[width=\linewidth]{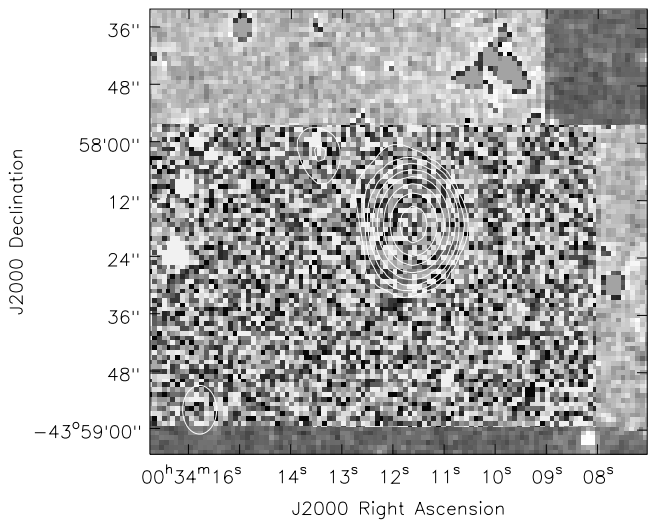}
\caption{Contour plot of the ATCA 20\,cm image of S427 superimposed on
the 3.6\um\ {\it Spitzer} image made as part of the SWIRE survey
(\citealt{Lonsdale2003}). Contours are drawn at 0.1\,mJy$\times$(1, 2,
4, ...) and the restoring beam was 10.3$\times$7.2\,arcsec. The
nearest infrared source, located towards the south-east, is more than
6\,arcsec away, making it very unlikely to be the infrared
counterpart. Source S433, visible as a two-contour object 24.6\,arcsec
north-east of S427, also was classified as an IFRS. It has a flux
density of 245\,\uJy.}
\label{fig:S427+ir}
\end{figure}

Here we present VLBI observations of four IFRS discovered in the
ATLAS/ELAIS field (\citealt{Middelberg2008a}). We selected S427 and
S509 because they were the strongest IFRS in the ATLAS/ELAIS field,
and S775 because it is very extended on arcsecond scales, showing
structures reminiscent of lobes and jets frequently seen in AGN. After
the observations it was discovered that the weak IFRS S433 was located
only 24.6\,arcsec north-east of S427 (Fig.~\ref{fig:S427+ir}), and was
well within the field of view of the VLBI array. The details of the
sources observed are listed in Table~\ref{tab:sources}.

\section{Observations}

\begin{figure}
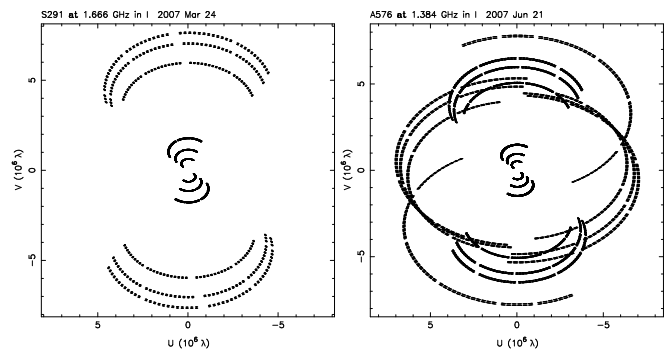

\centering
\includegraphics[width=0.5\linewidth, angle=270]{0454fig2.ps}
\includegraphics[width=0.5\linewidth, angle=270]{0454fig3.ps}
\caption{Plots showing the $(u,v)$ coverage of the observations of S427
on 24 March (left) and of S775 on 21 June (right).}
\label{fig:uvcov}
\end{figure}

We observed the IFRS with the Australian Long Baseline Array (LBA) in
phase-referencing mode. On 24 March 2007 the observing frequency was
1.6\,GHz and the participating antennas were the Australia Telescope
Compact Array (ATCA), the 64\,m Parkes telescope, the 22\,m Mopra
telescope and the 26\,m telescope near Hobart. On 21 June 2007 the
observing frequency was 1.4\,GHz and the same array was used, plus the
30\,m telescope at Ceduna. The ATCA, Parkes and Mopra telescopes
recorded a total bandwidth of 64\,MHz in both polarizations, using
2-bit sampling, subdivided in 16\,MHz wide IF channels. Both
right-hand and left-hand circular polarization were recorded. The
Hobart and Ceduna telescopes recorded a total bandwidth of 32\,MHz
with the same setup. Both observing runs lasted for 12\,h. The total
$(u,v)$ coverage of the two observing runs is shown in
Fig.~\ref{fig:uvcov}.

These are the first published LBA observations using Ceduna at
1.4\,GHz. The design of Ceduna (an old telecommunications dish donated
to the University of Tasmania in 1995) had not foreseen observations
below several GHz, and frequencies lower than 2\,GHz are blocked by
waveguides in the Nasmyth focus. Efforts are now underway to work
around this limitation by installing a tertiary mirror at the centre
of the main reflector. This bypasses the Nasmyth focus entirely and
allows the reception of frequencies below 2\,GHz. At the time of our
observations, a test system consisting of a fixed tertiary and an
uncooled 1.4\,GHz receiver had been installed, yielding a system
temperature of around 1260\,K in RCP and 1390\,K in LCP. In the near
future, a cooled receiver and moveable tertiary reflector will allow
Ceduna to routinely participate in VLBI observations at sky
frequencies between 1.2\,GHz and 1.8\,GHz, in addition to its existing
higher frequency capabilities.

On 24 March, the coordinates of IFRS S427 and S509 were observed for
5\,min each, followed by a 3\,min-scan on the nearby phase calibrator
0022-423, which has an arcsec-scale flux density (monitored with the
ATCA) of 2.82\,Jy. S433 was also in the field of view of these
observations. On 21 June, 8\,min-scans of the IFRS S775 were followed
by a 3\,min-scan of the same calibrator. One of the fringe finders
1921-293 and 0104-408 was observed every two hours in either run. The
LBA sensitivity
calculator\footnote{http://www.atnf.csiro.au/vlbi/calculator}
predicted a baseline sensitivity of 3\,mJy for a 3\,min calibrator
scan on the least sensitive baseline between Mopra and Hobart. The
predicted image sensitivities, using all data, were around 50\,\uJy\
for the targets.

\begin{table*}
\centering
\caption{Source parameters and results from our observations.  Column
1: source name; column 2: formal IAU designation, which for brevity we
do not use elsewhere in this paper; columns 3-4: coordinates used in
the observations. The target coordinates differ from those published
in \cite{Middelberg2008a} because they were taken from an early
version of the 1.4\,GHz ATCA image; column 5: ATCA 1.4\,GHz flux
density by \cite{Middelberg2008a} in mJy; column 6: integrated flux
density found in the VLBI images in mJy; column 7: peak flux density
of the VLBI images in mJy; column 8: rms noise measured from the VLBI
images in mJy. The peaks in the last three sources all are below
$5\,\sigma$, and the locations of the brightest pixel does not
coincide with the arcsec-scale source position. The VLBI flux
densities were all measured at 1.6\,GHz, with the exception of
those marked with a superscript asterisk, which were made at 1.4\,GHz.}
\label{tab:sources}
\centering
\begin{tabular}{llcccccc}
\hline
\hline
Source   & IAU designation             & RA                & Dec             & $S_{\rm 1.4\,GHz}$ & $S_{\rm VLBI}$ & $S_{\rm VLBI,max}$ & rms$_{\rm VLBI}$   \\
         &                             &                   &                 & mJy                & mJy            & mJy                & mJy   \\
\multicolumn{1}{c}{(1)}&
\multicolumn{1}{c}{(2)}&
\multicolumn{1}{c}{(3)}&
\multicolumn{1}{c}{(4)}&
\multicolumn{1}{c}{(5)}&
\multicolumn{1}{c}{(6)}&
\multicolumn{1}{c}{(7)}&
\multicolumn{1}{c}{(8)}\\
\hline
0022-423 & PKS 0022-423                & 00:24:42.989741   & -42:02:03.94796 & 2820               & 2690           & 1750               & 6.4   \\
         &                             &                   &                 &                    & 2390$^*$       & 1790$^*$           & 2.7$^*$\\
S427     & ATELAIS J003411.59-435817.0 & 00:34:11.59       & -43:58:17.036   & 21.4               & 12.5           & 8.2                & 0.14  \\
S509     & ATELAIS J003138.63-435220.8 & 00:31:38.64       & -43:52:20.824   & 22.2               & ...            & $<0.27$            & 0.065 \\
S433     & ATELAIS J003413.43-435802.4 & 00:34:13.43       & -43:58:02.470   & 0.2                & ...            & $<0.27$            & 0.069 \\
S775     & ATELAIS J003216.05-433329.6 & 00:32:16.01       & -43:33:37.092   & 3.6                & ...            & $<0.26^*$          & $0.055^*$ \\
\hline
\end{tabular}
\end{table*}

\section{Calibration}

\begin{figure}
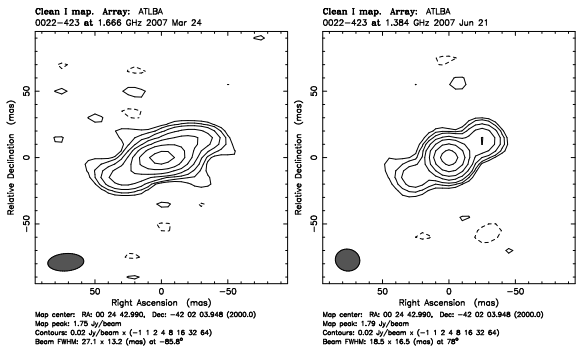

\centering
\includegraphics[width=0.5\linewidth, angle=270]{0454fig4.ps}
\includegraphics[width=0.5\linewidth, angle=270]{0454fig5.ps}
\caption{Contour plots of the phase calibrator 0022-423 observed on 24
March (left) and 21 June (right). Contours are drawn at
20\,mJy$\times$(-1, 1, 2, 4, ...) and the restoring beam was
27.1\,mas$\times$13.2\,mas (left) and 18.5\,mas$\times$16.5\,mas
(right). The observations on 21 June used a slightly improved position
for the calibrator.}
\label{fig:phasecal}
\end{figure}

The lower and upper 10 out of 64 channels in each IF had low
amplitudes and were flagged. Also, after source changes the correlator
produced data although the antennas had not yet arrived at the source
positions, requiring the flagging of 30\,s of the beginning of all
scans, and occasionally more. A short section of a fringe-finder
observation was fringe-fitted to obtain residual and instrumental
delays, and phase offsets for each IF and polarization
separately. These solutions were applied to the entire data set, and
allowed averaging across the band in the subsequent fringe-fitting of
the phase calibrator. The phase calibrator was detected on all
baselines throughout the experiments with a high signal-to-noise ratio
(SNR). Initial amplitude calibration was carried out using $T_{\rm
sys}$ values measured during the observations, and known antenna
gains. The phase calibrator was then imaged in Difmap
(\citealt{Shepherd1997}). Amplitude self-calibration was performed in
Difmap with a 720\,min solution interval, yielding small ($<$20\,\%)
corrections for the antenna gains. The complex gains from this
procedure were then applied to the targets. Contour plots of the
calibrator are shown in Fig.~\ref{fig:phasecal}. An area of $2''$
centred on the targets' arcsec-scale emission was imaged to look for
the sources.

\section{Results and discussion}

Only one of the three targets, S427, was detected with good SNR. The
other three fields were completely devoid of emission. In the case of
S433 the loss of sensitivity due to wide-field effects (bandwidth
smearing, time smearing, and primary beam attenuation) is of the order
of only a few percent on the longest baselines, and hence is
negligible. Therefore its non-detection is just as significant as the
other non-detections.

\begin{figure}
\centering
\includegraphics[width=\linewidth, angle=270]{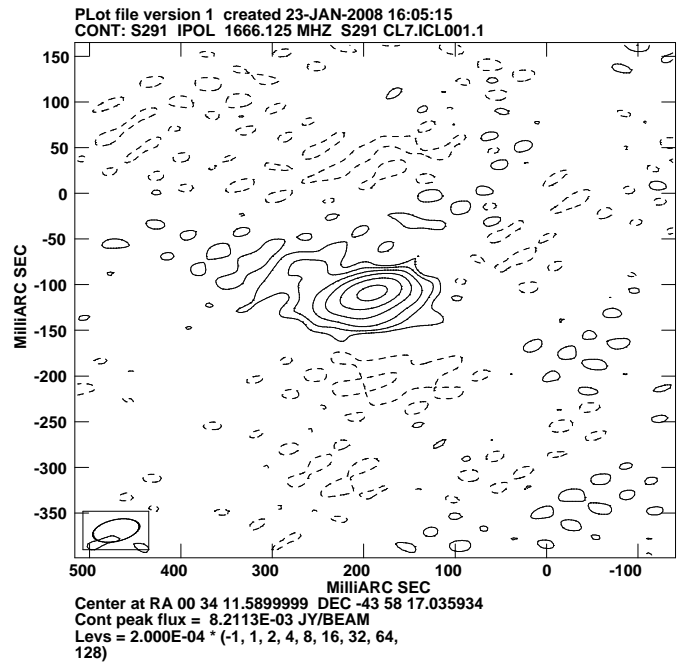}
\caption{Contour plot of the IFRS S427. Contours are drawn at
0.2\,mJy$\times$(-1, 1, 2, 4, ...) and the restoring beam was
51.7\,mas$\times$23.6\,mas.}
\label{fig:S427}
\end{figure}

\subsection{S427}

The image of S427 in Fig.~\ref{fig:S427} displays a slightly, though
significantly extended point source, with the highest sidelobes at a
level of 0.3\,mJy. Even though the extension is weak, it is unlikely
to be an imaging artefact because a) the image quality could be
improved considerably by including the extension in the model, and b)
because the flux density measured on the shortest baseline was about a
factor of two higher than that measured on the longest baseline. The
peak flux density in the image is 8.2\,mJy and the integrated flux
density is 12.5\,mJy. The limited $(u,v)$ coverage defied attempts to
develop a more detailed model, and caused the noise in regions away
from the source to be dominated by sidelobes rather than receiver
noise. On the longest baselines the source has a flux density of
7\,mJy, from which we infer a minimum brightness temperature of
$T_{\rm B,min}=3.6\times10^6$\,K, indicating non-thermal emission. We
therefore conclude that the source contains an AGN.

A VLBI image is a significant step towards understanding these
mysterious objects. First, we now know that the detection by
\cite{Norris2007b} is not a single, isolated event. Although VLBI
detections appear not to be the rule, they also do not appear to be
rare. Second, the extension visible in Fig.~\ref{fig:S427} allows
one to make statements about the source geometry and to compare it to
observations at lower resolution.

S427 has an arcsec-scale flux density of (21.4$\pm$1.07)\,mJy
(\citealt{Middelberg2008a}), 58\,\% of which we recovered on baselines
longer than 500\,k$\lambda$, or on scales smaller than 410\,mas, and
33\,\% of which are detected on baselines longer than 6.5\,M$\lambda$,
or on scales smaller than 32\,mas. 

\subsection{The spectrum of S427}

Following up on these results, on 27 and 29 April 2008 we observed
S427 with the ATCA at 4.8\,GHz and 8.6\,GHz, using director's time
during a maintenance period. At 4.8\,GHz we were able to combine data
from the 6A and 750A configurations whereas the 8.6\,GHz data were
only observed in the 750A configuration, and therefore yielded an
image with a lower resolution and image fidelity than the 4.8\,GHz
data. The bandwidth was 128\,MHz in two polarizations, and the
integration times were 11\,h at 4.8\,GHz and 5.6\,h at 8.6\,GHz.

The 4.8\,GHz image (not shown) displays a marginally resolved source
which is adequately described by a Gaussian with $B_{\rm
max}=5.7^{\prime\prime}$ and $B_{\rm min}=3.5^{\prime\prime}$ in ${\rm
PA}=-39^\circ$ (north through east). The deconvolved size was found to
be $1.6^{\prime\prime}\times0.7^{\prime\prime}$ in ${\rm
PA}=-63^\circ$, so the source is close to being unresolved. The
8.4\,GHz image (also not shown) displays a slightly extended source
with $4.2^{\prime\prime}\times3.6^{\prime\prime}$ in ${\rm
PA}=51^\circ$ and a deconvolved size of
$2.8^{\prime\prime}\times2.5^{\prime\prime}$ in ${\rm
PA}=58^\circ$. The integrated flux densities were found to be
$S_{4800}=(6.4\pm1.3)$\,mJy and $S_{8640}=(1.8\pm0.4)$\,mJy.

The PA of the deconvolved Gaussian at 8.6\,GHz is in approximate
agreement with the PA found at 1.4\,GHz, but this extension has not
been found at 4.8\,GHz. Our observations at 8.4\,GHz had significantly
lower SNR than those at 4.8\,GHz, which may have affected the
deconvolution. However, the SNR at 8.4\,GHz is too high to be the only
reason for this discrepancy, and therefore the deconvolved size of
S427 at this frequency is not understood.

From a 2.3\,GHz follow-up survey of the ATLAS/ELAIS field currently
being conducted by us with the ATCA, and from the SUMSS survey
(\citealt{Bock1999}) we have obtained two more spectral points for
S427. The catalogued flux density from the SUMSS survey at a frequency
of 843\,MHz and with a resolution of $45''\times64''$ was
(42.7$\pm$1.5)\,mJy, and from our preliminary 2.3\,GHz image at a
centre frequency of 2.424\,GHz and with resolution of $52''\times28''$
we have obtained a flux density of (12.5$\pm$1.9)\,mJy. We have
convolved the 1.4\,GHz ATLAS/ELAIS image (with a centre frequency
1.382\,GHz and a resolution of $10.3''\times7.2''$, but sufficient
short-baseline coverage for a low-resolution image) with an
appropriate Gaussian kernel to generate an image with a resolution
matching that at 2.3\,GHz, and found a flux density of 22.7\,mJy. This
is within the errors of the flux density measured from the
higher-resolution image and indicates that there is no faint emission
on scales of tens of arcsec, as would be expected from a very compact
source. The spectral indices therefore can be computed from the
full-resolution 1.4\,GHz image, and are $\alpha^{843}_{1.382}=-1.39$
and $\alpha^{1.382}_{2.424}=-0.96$.

\begin{figure}[htpb!]
\centering
\includegraphics[width=\linewidth]{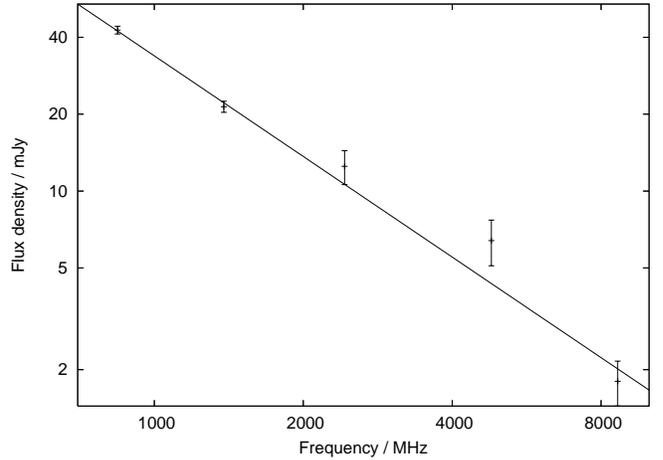}
\caption{The arcsec-scale radio spectrum of S427 between 843\,MHz and
8640\,MHz, and a power-law fitted to the data.}
\label{fig:spectrum}
\end{figure}

The five flux density measurements now available are shown in
Fig.~\ref{fig:spectrum}. We also show a power-law function fitted to
the data, which is able to represent the data rather well, and which
has an exponent of -1.31, showing that the source has a very steep
spectrum over a decade in frequency.

\subsection{Modelling the properties of S427 at two redshifts}

Following the discussion by \cite{Norris2007b}, we model the emission
of S427 assuming it has redshifts of $z=1$ and $z=7$ (the results of
which we give in brackets). The measured flux densities need to be
corrected for redshift ($k$-correction) and spectral index ($\alpha$,
$S\propto\nu^\alpha$). S427 has a 4.8\,GHz flux density of 6.4\,mJy
which, when corrected for redshift and spectral index, corresponds to
a rest-frame 5\,GHz luminosity of $7.9\times10^{25}$\,W\,Hz$^{-1}$
($5.5\times10^{28}$\,W\,Hz$^{-1}$ if z=7, using $H_0$=71, $\Omega_{\rm
M}=0.27$ and $\Omega_{\rm vac}=0.73$). At z=7 this is approximately
the luminosity of 3C\,273, hence is high (e.g.,
\citealt{O'Dowd2002}), but not impossible.

The size of the source can be estimated as follows. In the 1.4\,GHz
ATCA image the source was well represented with a Gaussian of
$10.62''\times7.89''$ in PA 174\degr. After deconvolution from the
restoring beam, the intrinsic source size was found to be
$3.30''\times2.74''$ in PA 95\degr. At z=1 the linear scale is
8.0\,kpc\,arcsec$^{-1}$ (5.3\,kpc\,arcsec$^{-1}$), and so the
intrinsic source size is of the order of 24\,kpc (16\,kpc).

We point out that the position angles of deconvolved sources in the
source catalogue by \cite{Middelberg2008a} are very evenly distributed
and do not appear to be biased in any direction. Hence the orientation
of S427, when deconvolved from the restoring beam in the arcsec-scale
image, is in agreement with the extension seen in the VLBI image.

It is surprising that S509 has spectral properties which are very
similar to S427, with $S_{2.424}=12.6$\,mJy, $S_{843}=37.6$\,mJy, and
thus $\alpha^{843}_{1.382}=-1.06$ and
$\alpha^{1.382}_{2.424}=-1.01$. Its intrinsic size at 1.4\,GHz after
deconvolution is $3.48''\times2.77''$ in PA 59\degr. Yet it remains
undetected in our VLBI observations which indicates that if it
contains an AGN it is comparatively weak or currently
inactive.

\subsection{The nature of S427}

Given its brightness temperature, S427 must contain an AGN, and this
AGN accounts for 58\,\% of its total flux density. Given its
compactness, it is not a radio lobe of a yet unidentified radio
galaxy, which was one of the possible explanations. Also, its steep
spectrum indicates that the radio emission from S427 is not dominated
by star formation, which is expected to produce a spectral index of
around -0.7.

S427 does not appear to be a ``standard'' FR\,I/II radio galaxy. Its
linear size of less than 24\,kpc if between redshifts of 1 and 7,
derived from the ATCA observations, is smaller by a factor of a few
than what is typically observed in FR\,I galaxies (although very few
are quite smaller than that), and its 1.4\,GHz luminosity of
log(L)=26.05\,W\,Hz$^{-1}$ if z=1 and log(L)=28.11\,W\,Hz$^{-1}$ if
z=7 is at the very high end of what is being observed
(\citealt{Owen1994}). FR\,II radio galaxies are even larger than
FR\,Is, but their luminosities can be as high as that of S427 if at
z=7. Furthermore, in FR\,I/II radio galaxies almost all of the
emission comes from the extended radio lobes, in particular at low
frequencies (due to the steep spectral index of the radio lobes). In
S427 33\,\% of the emission comes from scales smaller than 250\,pc,
which would be unusual for an FR\,II. We conclude that although we
cannot rule out that S427 is a FR\,I/II radio galaxy, its extent,
luminosity and internal distribution of emission make it unlikely.

A large fraction ($\sim$40\,\%, \citealt{Odea1998}) of radio sources
are the Compact Steep-Spectrum (CSS) and Gigaherz-Peaked Spectrum
(GPS) sources. CSS and GPS sources are very small yet strong radio
sources. They are contained within their host galaxies, with CSS
source being smaller than 20\,kpc, whereas GPS sources are smaller
than 1\,kpc, so that they are smaller than the narrow-line region. The
two competing models to explain their properties are that they are
either ``frustrated'' radio galaxies which are confined by their host
galaxy's very dense ISM, ot that they are young objects which
eventually will evolve into large radio galaxies. A detailed review
can be found in \cite{Odea1998}. Given its extent of more than 1\,kpc
and its power-law spectrum between 843\,MHz and 8640\,MHz, we can rule
out that S427 is a GPS source.

A general requirement for sources classified as CSS is a spectral
index of less than -0.5, a source size of less than about 20\,kpc, and
1.4\,GHz luminosities of more than $10^{25}$\,W\,m$^{-2}$\,Hz$^{-1}$
(\citealt{Odea1998}).  S427 has log($L_{1.4}$)=26.1 if z=1 and
log($L_{1.4}$)=28.1 if z=7. Its size and luminosity are therefore in
good agreement with typical CSS luminosities, if S427 is at high
redshifts.

Using VLBI observations of seven strong CSS, \cite{Tzioumis2002}
detect double-lobed structures in all objects. They also find that in
most of their objects more than 50\,\% of the arcsec-scale flux
density is contained in the VLBI images. However, in one case the
amount of emission resolved out by the VLBI observations is as high as
70\,\%. Within the limits of our observations, S427 does not display a
double-lobed structure on mas scales and so is not a typical CSS.

The infrared properties of CSS sources are not well studied, so it is
difficult to make a statement about whether S427's IR properties are
typical for a CSS or not. \cite{Heckman1994} find that the ratio of IR
to radio luminosity of radio galaxies is relatively independent of the
object class (FR\,II, quasar, GPS/CSS, and so on). This was confirmed
in similar studies carried out by \cite{Hes1995}, and by
\cite{Fanti2000}. However, these authors all studied predominantly 3C
sources which are selected to be very bright in the radio, and so
their samples are plagued by Malmquist bias and their conclusions may
not be applicable to IFRS. However, given the lack of other data we
estimate the IR luminosity of a CSS at high redshift.
\cite{Heckman1994} present a sample of {\it IRAS}-detected GPS/CSS
sources. Their 12\,\um\ flux densities typically are of the order of
tens of mJy, but only one object classified as CSS by
\cite{Odea1998}, 3C\,48, has been detected at wavelengths between
12\,\um\ and 100\,\um . 3C\,48 has a redshift of 0.367, and if shifted
to z=1 would have a (k-corrected) 24\,\um\ flux density of $S_{\rm
24\um}=10.6$\,mJy and a 3.6\,\um\ flux density of $S_{\rm
3.6\um}=0.5$\,mJy, which would have been detected in the SWIRE
survey. However, at z=7 it would appear as a source with $S_{\rm
24\um}=4.6\,\uJy$ and $S_{\rm 3.6\um}=0.9\,\uJy$, which is much too
faint for the SWIRE survey. Hence if S427's host galaxy is similar to
that of 3C48 it must be at high redshift.

\cite{Dicken2008} have recently carried out an investigation of the
radio and infrared properties of powerful, radio-loud radio galaxies
at intermediate redshifts using literature data, ATCA, VLA and {\it
Spitzer} observations. Their data suggest that the 24\,\um\ flux
densities of radio galaxies are typically one to two orders of
magnitude lower than the 1.4\,GHz flux densities, independent of
object type. Their study includes the three known CSS sources
PKS\,0252-71, PKS\,1151-34, PKS\,1814-63 and the GPS source
PKS\,1934-63. The radio and infrared properties of these sources
follow the trend of their sample. For S427 that implies that the
expected 24\,\um\ flux density would be between 0.2\,mJy and 2\,mJy
(compared to 160\,mJy \citealt{Heckman1994} find for 3C\,48). The
1\,$\sigma$ sensitivity of the 24\,\um\ ATLAS/{\it Spitzer}
observations is 252\,\uJy\ hence the non-detection of 24\,\um\
emission in the ATLAS survey is consistent with the observations by
\cite{Dicken2008}. However, we note that their work is concerned with very
bright radio sources, and that the conclusions made are unlikely to be
applicable to a poorly understood class of objects such as the IFRS.

\section{Conclusions}

We present the first VLBI image of an Infrared-Faint Radio Source and
report the non-detection of three more IFRS discovered in the ATCA
1.4\,GHz observations of the ATLAS/ELAIS field. The main result of our
observations is that S427 harbours an AGN, and that it is not simply a
radio lobe of an unidentified radio galaxy. The size, spectrum, and
radio and IR luminosity of the detected IFRS S427 are consistent with
those of a high-redshift Compact Steep-Spectrum source, and are
inconsistent with a standard L$_*$ galaxy at z$<$1, or a FR\,I/II
galaxy at any redshift.

Together with the two IFRS observed by Norris et al. (2007) the number
of IFRS observed with VLBI is now 6, and the number of detections is
2, showing that at least some fraction of IFRS are associated with
AGN. We plan further VLBI studies to determine the nature of these
enigmatic objects.

\bibliography{refs}

\begin{acknowledgements}
We thank the staff of the Australian Long Baseline Array for making
the observations presented in this paper possible, and we thank Brett
Reid for installing the 1.4\,GHz test feed on the Ceduna antenna.
\end{acknowledgements}

\end{document}